\documentclass[twocolumn,showpacs]
{revtex4}

\usepackage{latexsym}
\usepackage{bm}

\def\tr{{\rm tr}}
\def\ket#1{\mid~\!\!\!{#1}~\!\!\rangle}
\def\bra#1{\langle~\!\!{#1}~\!\!\!\mid}

\begin{document}

\title{Mixing Property of Quantum Relative Entropy}

\author{Fedor Herbut}
\affiliation {Serbian Academy of Sciences
and Arts, Knez Mihajlova 35, 11000
Belgrade, Serbia and Montenegro}

\email{fedorh@infosky.net}

\date{\today}

\begin{abstract}
An analogue of the mixing property of
quantum entropy is derived for quantum
relative entropy. It is applied to the
final state of ideal measurement and to
the spectral form of the second density
operator. Three cases of states on a
directed straight line of relative
entropy are discussed.
\end{abstract}

\pacs{03.65.Ta 03.67.-a} \maketitle

\rm Relative entropy plays a fundamental
role in quantum information theory (see
p. 15 in \cite{O-P} and the review
articles \cite{Vedral}, \cite{Schum},
which have relative entropy in the
title).

The {\it relative entropy}
$S(\rho||\sigma)$ of a state (density
operator) $\rho$ with respect to a state
$\sigma$ is by definition
$$S(\rho||\sigma)\equiv \tr [\rho log(\rho )]-\tr
[\rho log(\sigma)]\eqno{(1a)}$$
$$\mbox{if}\quad \mbox{supp}(\rho ) \subseteq
\mbox{supp}(\sigma );\eqno{(1b)}$$
$$\mbox{or else}\quad
S(\rho||\sigma)=+\infty \eqno{(1c)}$$
(see p. 16 in \cite{O-P}). By "support"
is meant the subspace that is the
topological closure of the range.

If $\sigma$ is singular and condition
(1b) is valid, then the orthocomplement
of the support (i. e., the null space) of
$\rho$, contains the null space of
$\sigma$, and both operators reduce in
supp$(\sigma )$. Relation (1b) is valid
in this subspace. Both density operators
reduce also in the null space of
$\sigma$. Here the $log$ is not defined,
but it comes after zero, and it is
generally understood that zero times an
undefined quantity is zero. We'll refer
to this as {\it the zero convention}.

The more familiar concept of (von
Neumann) quantum entropy, $S(\rho )\equiv
-\tr [\rho log(\rho )]$, also requires
the zero convention. If the state space
is infinite dimensional, then, in a
sense, entropy is almost always infinite
(cf p.241 in \cite{Wehrl}). In
finite-dimensional spaces, entropy is
always finite.

In contrast, relative entropy is often
infinite also in finite-dimensional
spaces (due to (1c)). Most results on
relative entropy with general validity
are {\it inequalities}, and the infinity
fits well in them. It is similar with
entropy. But there is one {\it equality
for entropy} that is much used, {\it the
mixing property} concerning {\it
orthogonal state decomposition} (cf p.
242 in \cite{Wehrl}):

$$\sigma =\sum_k w_k\sigma_k,\eqno{(2)}$$
$\forall k:\enskip w_k\geq 0$; for
$w_k>0$, $\sigma_k>0,\enskip \tr
\sigma_k=1$; $\sum_kw_k=1$. Then
$$S(\sigma )=H(w_k)+
\sum_kw_kS(\sigma_k),\eqno{(3a)}$$
$$H(w_k)\equiv
-\sum_k[w_klog(w_k)] \eqno{(3b)}$$ being
the Shannon entropy of the probability
distribution $\{w_k:\forall k\}$.

The {\it first aim} of this article is to
derive an analogue of (3a), which will be
called {mixing property of relative
entropy}. The {\it second aim} is to
apply it to the derivation of two
properties of the final state in ideal
measurement, and to the spectral
decomposition of $\sigma$ in the general
case.

We will find it convenient to make use of
an {\it extension} $log^e$ of the
logarithmic function to the entire real
axis:
$$\mbox{if}\quad 0<x:\qquad
log^e(x)\equiv log(x),\eqno{(4a)}$$ ,
$$\mbox{if}\quad
x\leq 0:\qquad log^e(x)\equiv 0.\quad
\eqno{(4b)}$$

The following elementary property of the
extended logarithm will be utilized.

Lemma 1: {\it If an orthogonal state
decomposition (2) is given, then
$$log^e(\sigma )
=\sum'_k [log(w_k)]Q_k+\sum'_k log^e
(\sigma_k),\eqno{(5)}$$ where $Q_k$ is
the projector onto the support of
$\sigma_k$, and the prim on the sum means
that the terms corresponding to $w_k=0$
are omitted.}

Proof: Spectral forms $\forall k, \enskip
w_k>0:\enskip
\sigma_k=\sum_{l_k}s_{l_k}\ket{l_k}
\bra{l_k}\quad$ (all $s_{l_k}$ positive)
give a spectral form $\sigma =
\sum_k\sum_{l_k}w_ks_{l_k}\ket{l_k}\bra{l_k}$
of $\sigma$ on account of the
orthogonality assumed in (2) and the zero
convention. Since numerical functions
define the corresponding operator
functions via spectral forms, one obtains
further
$$log^e(\sigma
)\equiv
\sum_k\sum_{l_k}[log^e(w_ks_{l_k})]\ket{l_k}
\bra{l_k}=$$
$$\sum_k'\sum_{l_k}[log(w_k)+log(s_{l_k})]
\ket{l_k} \bra{l_k}=$$
$$\sum_k'[log(w_k)]Q_k+\sum_k'
\sum_{l_k}[log(s_{l_k})]\ket{l_k}
\bra{l_k}.$$ (In the last step
$Q_k=\sum_{l_k}\ket{l_k}\bra{l_k}$ for
$w_k>0$ was made use of.) The same is
obtained from the RHS of (5) when the
spectral forms of $\sigma_k$ are
substituted in it. \hfill $\Box$

Now we come to the main result.

Theorem 1: {\it Let condition (1b) be
valid for the states $\rho$ and $\sigma$,
and let an orthogonal state decomposition
(2) be given. Then
$$S(\rho||\sigma)=S\Big(\sum_kQ_k\rho
Q_k\Big)-S(\rho )+$$
$$H(p_k||w_k)+\sum_kp_k S(Q_k\rho
Q_k/p_k||\sigma_k),\eqno{(6)}$$ where,
for $w_k>0$, $Q_k$ projects onto the
support of $\sigma_k$, and $Q_k\equiv 0$
if $w_k=0$, $p_k\equiv \tr (\rho Q_k)$,
and
$$H(p_k||w_k)\equiv
\sum_k[p_klog(p_k)]-\sum_k[p_klog(w_k)]
\eqno{(7)}$$ is the classical discrete
counterpart of the quantum relative
entropy, valid because $(p_k>0)\enskip
\Rightarrow (w_k>0)$.}

One should note that the claimed validity
of the classical analogue of (1b) is due
to the definitions of $p_k$ and $Q_k$.
Besides, (2) implies that $(\sum_kQ_k)$
projects onto supp$(\sigma )$. Further,
as a consequence of (1b),
$(\sum_kQ_k)\rho =\rho$. Hence, $\tr
\Big(\sum_kQ_k\rho Q_k\Big)=1$.

Proof of theorem 1: We define $$\forall
k,\enskip p_k>0:\quad \rho_k\equiv
Q_k\rho Q_k/p_k.\eqno{(8)}$$ First we
prove that (1b) implies $$\forall
k,\enskip p_k>0:\quad
\mbox{supp}(\rho_k)\subseteq \mbox{supp}
(\sigma_k).\eqno{(9)}$$

Let $k$, $p_k>0$, be an arbitrary fixed
value. We take a pure-state decomposition
$$\rho
=\sum_n\lambda_n\ket{\psi_n}\bra{\psi_n}
\eqno{(10a)},$$  $\forall n:\enskip
\lambda_n>0$. Applying $Q_k...Q_k$ to
(10a), one obtains another pure-state
decomposition
$$Q_k\rho Q_k=p_k\rho_k
=\sum_n\lambda_nQ_k\ket{\psi_n}\bra{\psi_n}
Q_k\eqno{(10b)}$$ (cf (8)). Let
$Q_k\ket{\psi_n}$ be a nonzero vector
appearing in (10b). Since (10a) implies
that $\ket{\psi_n}\in \mbox{supp}(\rho )$
(cf Appendix (ii)), condition (1b)
further implies $\ket{\psi_n}\in
\mbox{supp}(\sigma )$. Let us write down
a pure-state decomposition
$$\sigma =\sum_m
\lambda'_m\ket{\phi_m}\bra{\phi_m}
\eqno{(11)}$$ with $\ket{\phi_1}\equiv
\ket{\psi_n}$. (This can be done with
$\lambda'_1>0$ cf \cite{Hadji}.) Then,
applying $Q_k...Q_k$ to (11) and taking
into account (2), we obtain the
pure-state decomposition
$$Q_k\sigma Q_k=w_k\sigma_k=\sum_m
\lambda'_mQ_k\ket{\phi_m}\bra{\phi_m}
Q_k. \eqno{(11b)}$$ (Note that $w_k>0$
because $p_k>0$ by assumption.) Thus,
$Q_k\ket{\psi_n}=Q_k\ket{\phi_1}\in
\mbox{supp}(\sigma_k)$. This is valid for
any nonzero vector appearing in (10b),
and these span supp$(\rho_k)$ (cf
Appendix (ii)). Therefore, (9) is valid.

On account of (1b), the standard
logarithm can be replaced by the extended
one in definition (1a) of relative
entropy: $$ S(\rho ||\sigma
)=-S(\rho)-\tr [\rho log^e(\sigma )].$$
Substituting (2) on the RHS, and
utilizing (5), the relative entropy
$S(\rho ||\sigma )$ becomes
$$-S(\rho )-\tr \Big\{\rho
\Big[\sum_k'[log(w_k)]Q_k+\sum_k'[
log^e(\sigma_k)]\Big]\Big\}=$$
$$-S(\rho )-\sum_k'[p_klog(w_k)]-\sum_k'\tr
[\rho log^e(\sigma_k)].$$ Adding and
subtracting $H(p_k)$ (cf (3b)), replacing
$log^e(\sigma_k)$ by
$Q_k[log^e(\sigma_k)]Q_k$, and taking
into account (7) and (8), one further
obtains
$$S(\rho ||\sigma
)=-S(\rho )+H(p_k)+H(p_k||w_k)+$$
$$-\sum_k'p_k\tr [\rho_klog^e(\sigma_k)].$$
(The zero convention is valid for the
last term because the density operator
$Q_k\rho Q_k/p_k$ may not be defined.
Note that replacing $\sum_k$ by $\sum_k'$
in (7) does not change the LHS because
only $p_k=0$ terms are omitted.)

Adding and subtracting the entropies
$S(\rho_k)$ in the sum, one further has
$$S(\rho ||\sigma
)=-S(\rho )+H(p_k)+H(p_k||w_k)+$$
$$\sum_k'p_kS(\rho_k)+\sum_k'p_k\{-S(\rho_k) -\tr
[\rho_klog^e(\sigma_k)]\}.$$ Utilizing
the mixing property of entropy (3a), one
can put $S\Big(\sum_kp_k\rho_k\Big)$
instead of
$[H(p_k)+\sum_k'p_kS(\rho_k)]$. Owing to
(9), we can replace $log^e$ by the
standard logarithm and thus obtain
the RHS(6). \hfill $\Box$\\

{\it Some Applications of the Mixing
Property} - Let $\rho$ be a state and
$A=\sum_ia_iP_i+\sum_ja_jP_j$ a spectral
form of a discrete observable (Hermitian
operator) $A$, where the eigenvalues
$a_i$ and $a_j$ are all distinct. The
index $i$ enumerates all the detectable
eigenvalues, i. e., $\forall i:\enskip
\tr (\rho P_i)>0$, and $\tr [\rho
(\sum_iP_i)]=1$.

After an {\it ideal measurement} of $A$
in $\rho$, the entire ensemble is
described by the {\it L\"{u}ders state}:
$$\rho_L(A)\equiv \sum_iP_i\rho
P_i\eqno{(12)}$$ (cf \cite{Lud}). (One
can take more general observables that
are ideally measurable in $\rho$ cf
\cite{Roleof}. For simplicity we confine
ourselves to discrete ones.)

Corollary 1: {\it The relative-entropic
"distance" from any quantum state to its
L\"{u}ders state is the difference
between the corresponding quantum
entropies:} $$S\Big(\rho ||\sum_iP_i\rho
P_i\Big)=S\Big(\sum_iP_i\rho
P_i\Big)-S(\rho ).\eqno{(13)}$$

Proof: First we must prove that
$$\mbox{supp}(\rho )\subseteq
\mbox{supp}\Big(\sum_iP_i\rho
P_i\Big).\eqno{(14)}$$ To this purpose,
we write down a decomposition (10a) of
$\rho$ into pure states. One has
$\mbox{supp}(\sum_iP_i)\supseteq
\mbox{supp}(\rho )$ (equivalent to the
certainty of $(\sum_iP_i)$ in $\rho$, cf
\cite{Roleof}), and the decomposition
(10a) implies that each $\ket{\psi_n}$
belongs to $\mbox{supp}(\rho )$. Hence,
$\ket{\psi_n}\in \mbox{supp}(\sum_iP_i)$;
equivalently,
$\ket{\psi_n}=(\sum_iP_i)\ket{\psi_n}$.
Therefore, one can write
$$\forall n:\quad \ket{\psi_n}=\sum_i(P_i
\ket{\psi_n}).\eqno{(15a)}$$ Further,
(10a) implies
$$\sum_iP_i\rho
P_i=\sum_i\sum_n\lambda_nP_i\ket{\psi_n}
\bra{\psi_n}P_i.\eqno{(15b)}$$ As seen
from (15b), all vectors
$(P_i\ket{\psi_n})$ belong to
supp$(\sum_iP_i\rho P_i)$. Hence, so do
all $\ket{\psi_n}$ (due to (15a)). Since
$\rho$ is the mixture (10a) of the
$\ket{\psi_n}$, the latter span
$\mbox{supp}(\rho )$. Thus, finally, also
(14) follows.

In our case $\sigma \equiv \sum_iP_i\rho
P_i$ in (6). We replace $k$ by $i$. Next,
we establish
$$\forall i:\quad Q_i\rho Q_i=P_i\rho
P_i.\eqno{(16)}$$ Since $Q_i$ is, by
definition, the support projector of
$(P_i\rho P_i)$, and $P_i(P_i\rho
P_i)=(P_i\rho P_i)$, one has $P_iQ_i=Q_i$
(see Appendix (i)). One can write
$P_i\rho P_i=Q_i( P_i\rho P_i)Q_i$, from
which then (16) follows.

Realizing that $w_i\equiv \tr (Q_i\rho
Q_i)=\tr (P_i\rho P_i)\equiv p_i$ due to
(16), one obtains $H(p_i||w_i)=0$ and
$$\forall i:\quad S(Q_i\rho Q_i/p_i ||P_i\rho
P_i/w_i)=0$$ in (6) for the case at
issue. This completes the proof.\hfill
$\Box$

Now we turn to a peculiar further
implication of corollary 1.

Let $B=\sum_k\sum_{l_k}b_{kl_k}P_{kl_k}$
be a spectral form of a discrete
observable (Hermitian operator) $B$ such
that all eigenvalues $b_{kl_k}$ are
distinct. Besides, let $B$ be more
complete than $A$ or, synonymously, a
refinement of the latter. This, by
definition means that
$$\forall k:\quad
P_k=\sum_{l_k}P_{kl_k}\eqno{(17)}$$ is
valid. Here $k$ enumerates both the $i$
and the $j$ index values in the spectral
form of $A$.

Let $\rho_L(A)$ and $\rho_L(B)$ be the
L\"{u}ders states (12) of $\rho$ with
respect to $A$ and $B$ respectively.

Corollary 2: {\it The states $\rho$,
$\rho_L(A)$, and $\rho_L(B)$ lie on a
straight line with respect to relative
entropy, i. e. $$S\Big(\rho ||
\rho_L(B)\Big)=S\Big(\rho
||\rho_L(A)\Big)+S\Big(\rho_L(A))||
\rho_L(B)\Big),\eqno{(18a)}$$ or
explicitly:} $$S\Big(\rho
||\sum_i\sum_{l_i}(P_{il_i}\rho
P_{il_i})\Big)=S\Big(\rho
||\sum_i(P_i\rho P_i)\Big)+$$
$$ S\Big(\sum_i(P_i\rho P_i)||
\sum_i\sum_{l_i}(P_{il_i} \rho
P_{il_i})\Big).\eqno{(18b)}$$

Note that all eigenvalues $b_{kl_k}$ of
$B$ with indices others than $il_i$ are
undetectable in $\rho$.

Proof follows immediately from corollary
1 because
$$S\Big(\rho ||\rho_L(B)\Big)
=\Big[S\Big(\rho_L(B)\Big)-
S\Big(\rho_L(A)\Big)\Big]+$$
$$\Big[S\Big(\rho_L(A)\Big)-S(\rho
)\Big],$$ and, as easily seen from (12),
$\rho_L(B)= \Big(\rho_L(A)\Big)_L(B)$ due
to $P_{il_i}P_{i'}=\delta_{i,i'}P_{il_i}$
(cf (17)).\hfill $\Box$

Next, we derive another consequence of
theorem 1.

Corollary 3: {\it Let $\{p_k:\forall k\}$
and $\{w_k:\forall k\}$ be probability
distributions such that $p_k>0\enskip
\Rightarrow \enskip w_k>0$. Then,
$$H(p_k||w_k)=S\Big(\sum_kp_k\ket{k}\bra{k}
||\sum_kw_k\ket{k}\bra{k}\Big),\eqno{(19)}$$
where the LHS is given by (7), and the
orthonormal set of vectors
$\{\ket{k}:\forall k\}$ is arbitrary.}

Proof: Applying (6) to the RHS of (19),
one obtains
$$RHS(19)=S\Big(\sum_kp_k\ket{k}\bra{k}\Big)
-S\Big(\sum_kp_k\ket{k}\bra{k}\Big)+$$
$$H(p_k||w_k)+\sum_kp_kS(\ket{k}\bra{k}||
\ket{k}\bra{k})=LHS(19).$$\hfill $\Box$

Finally, a quite different general result
also follows from the mixing property
(6).

Theorem 2: {\it Let $S(\rho ||\sigma )$
be the relative entropy of any two states
such that (1b) is satisfied. Let,
further,
$$\sigma
=\sum_kw_k\ket{k}\bra{k}\eqno{(20)}$$ be
a spectral form of $\sigma$ in terms of
eigenvectors. Then $$S(\rho ||\sigma )=
S\Big(\rho ||\sum_k(\ket{k}\bra{k}\rho
\ket{k}\bra{k})\Big)+$$
$$S\Big(\sum_k(\ket{k}\bra{k} \rho
\ket{k}\bra{k})||\sigma
\Big).\eqno{(21)}$$ Thus, the states
$\rho$, $\sum_k(\ket{k}\bra{k} \rho
\ket{k}\bra{k})$ (cf (20) for
$\ket{k}\bra{k}$), and $\sigma$ lie on a
directed straight line of relative
entropy.}

Proof: Application of (6) to the LHS(21),
in view of (20), leads to
$$S(\rho ||\sigma )=S\Big(\sum_k(\ket{k}\bra{k}
\rho \ket{k}\bra{k})\Big)-S(\rho )+$$
$$H(p_k||w_k)+\sum_kp_kS(\ket{k}\bra{k}
||\ket{k}\bra{k}).$$ In view of $\enskip
p_k= \bra{k}\rho \ket{k}$, (13), (19),
and (20), this equals RHS(21).\hfill
$\Box$

It is well known that the
relative-entropic "distance", unlike the
Hilbert-Schmidt (HS) one, fails to
satisfy the triangle rule, which requires
that the distance between two states must
not exceed the sum of distances if a
third state is interpolated. But, and
this is part of the triangle rule, one
has equality if and only if the
interpolated state lies on a straight
line with the two states. As it is seen
from corollary 2 and theorem 2 as
examples, the relative-entropic
"distance" does satisfy the equality part
of the triangle rule.

An interpolated state lies on the HS line
between two states if and only if it is a
convex combination of the latter.
Evidently, this is not true in the case
of relative entropy.

Partovi \cite{Partovi} has recently
considered three states on a directed
relative-entropic line: a general
multipartite state $\rho_1 \equiv
\rho_{AB\dots N}$, a suitable separable
multipartite state $\rho_2\equiv
\rho_{AB\dots N}^S$ with the same
reductions $\rho_A, \rho_B, \dots
,\rho_N$, and finally $\rho_3\equiv
\rho_A\otimes \rho_B\otimes \dots \otimes
\rho_N$. The mutual information in
$\rho_1$ is taken to be its total
correlations information. It is well
known that it can be written as the
relative entropy of $\rho_1$ relative to
$\rho_3$. The straight line implies:
$$S(\rho_1||\rho_2)=
S(\rho_1||\rho_3)-S(\rho_2||\rho_3).$$ To
my understanding, it is Partovi's idea
that if $\rho_2$ is as close to $\rho_1$
as possible (but being on the straight
line and having the same reductions),
then its von Neumann mutual information
$S(\rho_2||\rho_3)$ equals the classical
information in $\rho_1$, and
$S(\rho_1||\rho_2)$ is the amount of
entanglement or quantum correlation
information in $\rho_1$.

Partovi's approach utilizes the
relative-entropy "distance" in the only
way how it is a distance: on a straight
line. One wonders why should the relative
entropy "distance" be relevant outside a
straight line, where it is no distance at
all cf \cite{V-P}, \cite{Plenio}. On the
other hand, these approaches have the
very desirable property of being
entenglement monotones. But so are many
others (see {\it ibid.}).

To sum up, we have derived the mixing
property of relative entropy $S(\rho
||\sigma )$ for the case when (1b) is
valid (theorem 1), and two more general
equalities of relative entropies
(corollary 3 and theorem 2), which follow
from it. Besides, two properties of
L\"{u}ders states (12) have been obtained
(corollary 1 and corollary 2). The mixing
property is applicable to any orthogonal
state decomposition (2) of $\sigma$.
Hence, one can expect a versatility of
its applications in quantum information
theory.\\

{\it Appendix} - Let $\rho
=\sum_n\lambda_n\ket{n}\bra{n}$ be an
arbitrary decomposition of a density
operator into ray projectors, and let $E$
be any projector. Then $$E\rho =\rho
\quad \Leftrightarrow \quad \forall
n:\enskip E\ket{n}=\ket{n}\eqno{(A.1)}$$
(cf Lemma A.1. and A.2. in
\cite{FHJP94}).

(i) If the above decomposition is an
eigendecomposition with positive weights,
then $\sum_n\ket{n}\bra{n}=Q$, $Q$ being
now the support projector of $\rho$, and,
on account of (A.1),
$$E\rho =\rho \quad \Rightarrow \quad
EQ=Q.\eqno{(A.2)}$$.

(ii) Since one can always write $Q\rho
=\rho$, (A.1) implies that all $\ket{n}$
in the arbitrary decomposition belong to
supp$(\rho )$. Further, defining a
projector $F$ so that supp$(F)\equiv$
span$(\{\ket{n}:\forall n\})$, one has
$FQ=F$. Equivalence (A.1) implies $F\rho
=\rho$. Hence, (A.2) gives $QF=Q$.
Altogether, $F=Q$, i. e., the unit
vectors $\{\ket{n}:\forall n\}$
span supp$(\rho)$.\\

\end{document}